\documentclass{ws-ijmpe}
\def\la{\langle}
\def\ra{\rangle}

\def\k2av{\la k_T^2\ra}

\begin{document}

\markboth{G.G.\,Barnaf\"oldi et al.}{
Does the Cronin Peak Disappear at LHC Energies?}
\catchline{}{}{}{}{}

\title{DOES THE CRONIN PEAK DISAPPEAR AT LHC ENERGIES? }

\author{\footnotesize GERGELY G\'ABOR BARNAF\"OLDI}
\author{P\'ETER L\'EVAI}
\address{RMKI Research Institute for Particle and Nuclear Physics, \\
P.O. Box 49, Budapest 1525, Hungary
bgergely@rmki.kfki.hu}

\author{GEORGE FAI}
\address{Center for Nuclear Research, Department of Physics, \\
Kent State University, Kent, OH 44242, USA\\
}

\author{G\'ABOR PAPP}
\address{Department of Theoretical Physics, E\"otv\"os University,\\
P\'azm\'any P. 1/A, Budapest 1117, Hungary\\
}

\author{BRIAN A. COLE}
\address{Nevis Laboratory, Columbia University,\\
New York, NY, USA\\
}

\maketitle

\begin{history}
\received{(received date)}
\revised{(revised date)}
\end{history}

\begin{abstract}
In this work we compare the nuclear modification factors in proton 
(deuteron) -- nucleus collisions at CERN SPS, FNAL and RHIC energies 
in a wide $p_T$ range. In these experiments the nuclear modification 
factor has shown an enhancement at $p_T \approx 4$~GeV/c. The height 
of this ``Cronin peak'' depends on the c.m. energy of the collision, 
as it is subject to stronger shadowing at higher energies.
One of the aims of this contribution is to analyze the shadowing 
phenomenon at lower ($2$ GeV/c $ \lesssim p_T \lesssim 4$ GeV/c) and 
intermediate ($4$ GeV/c $ \lesssim p_T \lesssim 8$ GeV/c) transverse 
momentum. Different shadowing parameterizations are considered and the 
obtained Cronin peaks are investigated at RHIC and LHC energies. 
\end{abstract}

\section{Introduction}

Enhancement of the hadron spectra in nuclear collisions is a strong 
nuclear effect. This was discovered in $pBe$, $pTi$ and $pW$ 
collisions and named after J.W.~Cronin~\cite{Cron75,Antr79} at FNAL. 
The measured enhancement is a $ \sim 40 \%$ in the lower and 
intermediate transverse momentum region ($2$ GeV/c 
$ \lesssim  p_T \lesssim 8$ GeV/c). Relativistic Heavy Ion Collider 
(RHIC) experiments measured a smaller ($\sim 10\%$) Cronin peak at 
higher energy, $\sqrt{s_{NN}}=200$ GeV in $dAu$ 
collisions~\cite{PHENIXdAu,PHENIXdAu05,STARdAu}. It is natural to ask 
the question: how will this effect appear at the energies of the 
forthcoming measurements at the Large Hadron Collider (LHC)? 

In this paper we are presenting our predictions for the pion spectra 
in proton-proton collisions as a reference at $\sqrt{s_{NN}}=900$
GeV and $8.8$ TeV energies. This leads us to predict the nuclear 
modification factor in $dPb$ collisions, taking into account initial 
state nuclear effects.

\section{Effects on Inclusive Pion Spectra in $pp$ Collisions}

A comparison of inclusive spectra and hadron-hadron correlations from 
$pp$ collisions to results of pQCD calculations shows that in this 
framework intrinsic transverse momentum ($k_T$) is necessary for the 
precise description of the 
data\cite{Wong98,Wang01,Yi02,Bp02,PHENIXcorr,JanRak,LevaiFai}. 
\begin{figure}[th]
\centerline{\rotatebox{0}{\psfig{file=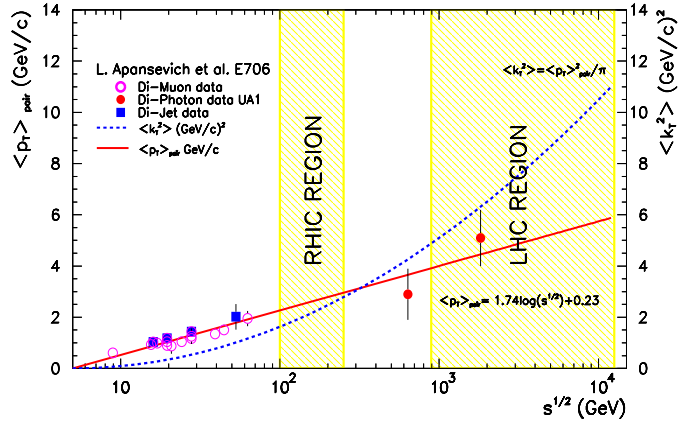,width=9cm,height=6.5cm}}}
\vspace*{8pt}
\caption{Estimation for the $\langle p_T \rangle_{pair} $ and for 
$ \langle k_T^2 \rangle $ value at different c.m. energies.} 
\label{fig:1}
\end{figure}

Several experiments (e.g. PHENIX\cite{PHENIXcorr,JanRak}, E706\cite{e706}) 
have measured the intrinsic transverse momentum using a produced 
hadron pair as a function of c.m. energy, as summarized in
Ref.\cite{zielinski}. It was found that 
$\langle p_T \rangle _{pair} \sim \log (\sqrt{s})$. These experimental
data and the linear fit is plotted on Fig.~\ref{fig:1}  
({\sl solid line}). The fitted linear function can be parameterized in 
the following form: 
\begin{equation}
\langle p_T \rangle _{pair} = \left( 1.74 \pm 0.12 \right) \cdot 
\log _{10} \left( \sqrt{s} \right) + \left(1.23 \pm 0.2 \right).
\label{fitkt}
\end{equation}
Calculations of $\langle k_T^2 \rangle$ are also shown ({\sl dashed 
curve}). The intrinsic $k_T$ was converted to the transverse momentum 
of the pair via 
$\langle k_T^2 \rangle_{pp} = \langle p_T \rangle ^2 _{pair} / \pi$.   

\begin{figure}[th]
\centerline{\psfig{file=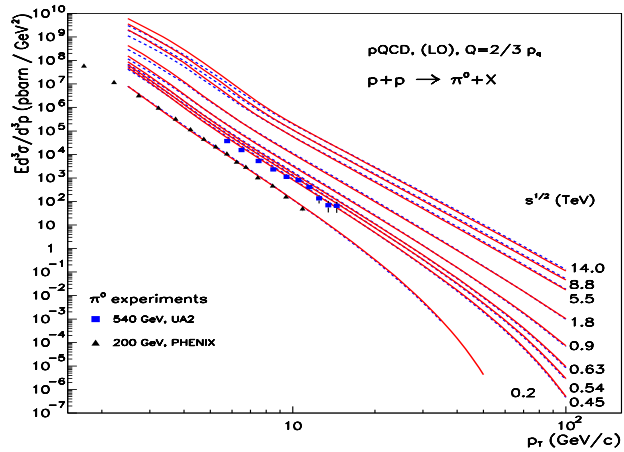,width=9cm,height=6.5cm}}
\vspace*{8pt}
\caption{Pion spectra in $pp$ collisions at 
different c.m. energies from RHIC to LHC energies, applying energy 
dependent $\langle k_T^2 \rangle $ (see. Fig~\ref{fig:1}). } 
\label{fig:2}
\end{figure}

While there are only a few experimental data points beyond RHIC 
energies, we can apply estimate (\ref{fitkt}) in calculations of pion 
spectra, using the appropriate intrinsic-$k_T$ values at given c.m. 
energy. The calculated $\pi^0$ spectra in $pp$ collisions are 
displayed in Fig.~\ref{fig:2}, where we compare this to experimental 
data from Refs.\cite{JanRak,ua2}. Our calculations use the 
GRV\cite{GRV} and HKN\cite{Shad_HKM,Shad_HKN} parton distribution functions. 

\begin{figure}[th]
\centerline{\psfig{file=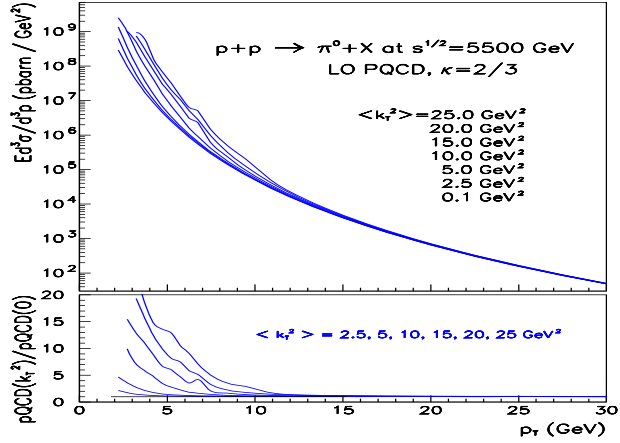,width=9cm,height=6.5cm}}
\vspace*{8pt}
\caption{The effect of the intrinsic $k_T$ in pion production
in $pp$ collisions at $\sqrt{s_{NN}}=5.5$ TeV. {\sl Upper panel} shows 
the spectra and the {\sl lower panel} presents the ratio relative to 
the zero intrinsic-$k_T$ calculation.} 
\label{fig:3}
\end{figure}

On Fig~\ref{fig:3} we are presenting the ``evolution'' of the pion 
spectra with different $ \langle k_T^2 \rangle $ values in $pp$ 
collisions at $5.5$ TeV c.m. energy. This shows that an increasing
intrinsic $k_T$ results in an enhancement of the pion spectra in 
the lower momentum region relative to the $\langle k_T^2 \rangle =0$ 
case. 

Considering a $\langle k_T^2 \rangle \sim 10-15$ 
GeV\textsuperscript{2}/c\textsuperscript{2} value at 
$\sqrt{s_{NN}}=5.5$ TeV c.m. energy, the modification is a factor of 
$\sim 5$ at the momentum region, $ p_T \sim 5$ GeV/c. Fig~\ref{fig:3} 
suggests that the effect of the intrinsic transverse momentum can be 
reasonably large at these high energies. This is interesting in itself, 
but it is even more important when we use calculated $pp$ hadron 
spectra in the nuclear modification factor. 

\section{Predictions for Pion Spectra in $dPb$ Collisions}

Pion production is calculated for $dPb$ collisions in a pQCD-improved 
parton model, described in Refs.\cite{Yi02,Bp02}. Within 
this framework we are taking into account the effect on intrinsic
transverse momenta in two aspects:
\begin{romanlist}[(ii)]
\item We are using a simple
generalization of the one- dimensional parton distribution functions
into $3$ dimensions, using a factorized form, 
\begin{equation}
f_{a/p}(x_a,{\bf k}_{Ta},Q^2) \,\,\,\, = \,\,\,\, f_{a/p}(x_a,Q^2) 
\cdot g_{a/p} ({\bf k}_{Ta}) \ , 
\end{equation}
where the function $f_{a/p}(x_a,Q^2)$ represents the standard
longitudinal  PDF as a function of $x_a$ at the factorization 
scale $Q$.  In the present calculation we use the GRV\cite{GRV} or 
HKN\cite{Shad_HKM,Shad_HKN} parameterizations. The partonic transverse-momentum 
distribution in two dimensions, $g_{a/p}({\bf k}_T)$, is assumed to be 
a Gaussian, characterized by the width $\langle k_T^2 \rangle$, 
sometimes referred to as the intrinsic-$k_T$ parameter.
\item Nuclear multiscattering is accounted for through a broadening 
of the incoming parton's transverse momentum distribution function, 
namely an increase in the width of the Gaussian:
\begin{equation}
\label{ktbroadpA}
\k2av_{pA} = \k2av_{pp} + C \cdot h_{pA}(b) \ .
\end{equation}
Here, $\k2av_{pp}$ is the width of the transverse momentum 
distribution of partons in $pp$ collisions\cite{Yi02,Bp02,Levai0306},
$h_{pA}(b)$ describes the number of {\it effective} $NN$ collisions
at impact parameter $b$, which impart an average transverse momentum
squared~$C$. The effectivity function $h_{pA}(b)$ can be written in
terms of the number of collisions suffered by the incoming proton in
the target nucleus. In Ref.\cite{Yi02} we have found  a limited
number of semi-hard collisions, $3 \leq \nu_{m} \leq 4$ and the value
$C = 0.35$ GeV$^2$/c$^2$.
\end{romanlist}

We calculate the nuclear modification factor as a function of c.m. 
energy. With increasing intrinsic $k_T$ the Cronin peak is found to 
shift towards higher $p_T$ values. At $\sqrt{s}=200$ GeV c.m. energy, 
the maximum of the Cronin peak was located at $p_T \approx 4$ GeV/c; 
at the LHC we expect this peak at $p_T \approx 5$ GeV/c and $8$ GeV/c, 
respectively, at $\sqrt{s}=900$ GeV and $8.8$ TeV c.m. energies.
This effect mirrors the recently measured experimental data by the
WA98\cite{wa98} and NA49\cite{na49} collaborations, where the Cronin 
peak seems to be at lower $p_T$ ($p_T \approx 2-3$ GeV/c) at 
$\sqrt{s}=17.3$ GeV c.m. energy.  

\begin{figure}[th]
\centerline{\psfig{file=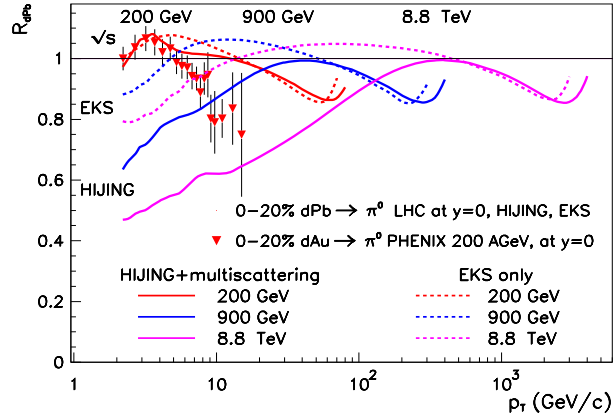,width=9cm,height=6.5cm}}
\vspace*{8pt}
\caption{The nuclear modification factor for pion production
in $0-10\%$ most central $dPb$ collisions at different c.m. energies.
Calculations have been performed with EKS ({\sl dashed lines}) and 
HIJING ({\sl solid lines}) shadowing parameterizations.} 
\label{fig:4}
\end{figure}

On Fig.~\ref{fig:4} we plot the $R_{dPb}$ nuclear modification factor
for the cases of the above mentioned two energies. For comparison we 
also show the latest experimental data by the 
PHENIX\cite{PHENIXdAu,PHENIXdAu05} on $R_{dAu}$ at $\sqrt{s}=200$ GeV.
We calculated the nuclear modification factor applying the 
$\sqrt{s}$-dependent intrinsic $k_T$ based on eq.~(\ref{fitkt}). 

In order to develop a feeling for the uncertainty of the shadowing 
parameterizations, we carried out our calculations with two 
shadowing parameterizations, EKS\cite{Shad_EKS} ({\sl dashed lines})
and HIJING\cite{Shad_HIJ,Shadxnw_uj} ({\sl solid lines})\footnote{The EKS 
parameterization contains some enhancement by definition 
as an anti-shadowing. Thus no multiple scattering was taken into 
account in this case.}. As Fig.~\ref{fig:4} shows, at lower $p_T$
the uncertainty is growing with increasing $\sqrt{s}$, corresponding 
to the limited information on shadowing with decreasing $x$. 
Close to $x\sim 1$ another ambiguity can be seen at the EMC 
region\cite{emc95,brian_emc}.

The shifting Cronin peak is suppressed by the strong shadowing
which is $\sim 40 \%$ for HIJING and $\sim 20\%$ for EKS. A 
remaining small `bump' is seen on the solid curves (HIJING 
parameterization). 

\section{Conclusions}

Based on experimental data we developed an approximation
for the values of $\k2av_{pp}$ at various c.m. energies. 
We have shown that the intrinsic $k_T$ has a non-negligible effect on 
inclusive pion production in proton-proton collisions at LHC energies. 
The increasing intrinsic $k_T$ causes a slight shift of the Cronin peak 
toward higher $p_T$ values. However, at LHC energies the strong 
$\sim 20-40 \%$  shadowing suppresses the Cronin peak. Overall, we 
expect that the Cronin peak will be totally suppressed at 
LHC energies in $dPb$ collisions by initial state effects. 

It is important to note on the other hand that, as it was 
pointed out in Refs.\cite{brian_emc}, at these high c.m. 
energies final state effects can also play a role, 
yielding more suppression in the nuclear modification factor.

\section*{Acknowledgments}

Our work was 
supported in part by Hungarian OTKA T043455, T047050, and NK62044,
by the U.S. Department of Energy under grant U.S. DE-FG02-86ER40251,
and jointly by the U.S. and Hungary under MTA-NSF-OTKA OISE-0435701.


\end{document}